\documentclass[acmsmall]{acmart}
\AtBeginDocument{%
  }

\usepackage{microtype}
\usepackage{graphicx}
\usepackage{subcaption}
\usepackage{xspace}
\usepackage{booktabs} 

\usepackage{hyperref}

\usepackage{amsmath}
\usepackage{amssymb}
\usepackage{mathtools}
\usepackage{amsthm}

\DeclareMathOperator*{\argmin}{arg\,min}

\usepackage{enumitem}

\usepackage{pifont}

\usepackage{graphicx}
\newcommand{\llmicon}[2]{%
  \raisebox{-0.15em}{\includegraphics[height=#1]{#2}}%
}

\usepackage{listings}
\usepackage[most]{tcolorbox}
\tcbuselibrary{breakable, skins, listings}

\usepackage{xcolor}

\definecolor{searchpurple}{RGB}{120,70,180}
\definecolor{plangreen}{RGB}{40,150,90}
\definecolor{gray}{gray}{0.4}
\usepackage{hyperref}

\usepackage{booktabs, multirow}
\usepackage{array}
\usepackage{colortbl}
\usepackage{wrapfig}
\definecolor{orangered}{RGB}{230,90,30}

\newcommand{\tool}{\textsc{IssueExec}\xspace}

\usepackage[capitalize,noabbrev]{cleveref}

\theoremstyle{plain}

\theoremstyle{definition}

\theoremstyle{remark}

\setcopyright{cc}
\setcctype{by}
\acmDOI{10.1145/3832290}
\acmYear{2026}
\acmJournal{PACMSE}
\acmVolume{3}
\acmNumber{ISSTA}
\acmArticle{ISSTA199}
\acmMonth{10}
\acmSubmissionID{issta26main-p2497-p}
\received{2026-01-30}
\received[accepted]{2026-04-16}

\begin{document}

\title{\tool: A Test-Driven Approach for Localizing Software Engineering Issues}


\author{Jiawei Liu}
\orcid{0009-0001-9823-0538}
\affiliation{%
  \institution{Shanghai Jiao Tong University}
  \city{Shanghai}
  \country{China}
}
\affiliation{%
  \institution{Shanghai Innovation Institute}
  \city{Shanghai}
  \country{China}
}
\email{amberwabi2003@sjtu.edu.cn}

\author{Yun Lin}
\orcid{0000-0001-8255-0118}
\affiliation{%
  \institution{Shanghai Jiao Tong University}
  \city{Shanghai}
  \country{China}
}
\email{lin\_yun@sjtu.edu.cn}

\author{Chenyan Liu}
\orcid{0009-0005-0554-4028}
\affiliation{%
  \institution{National University of Singapore}
  \city{Singapore}
  \country{Singapore}
}
\email{chenyan@u.nus.edu}

\author{Yu Qian}
\orcid{0009-0008-0745-0350}
\affiliation{%
  \institution{Shanghai Jiao Tong University}
  \city{Shanghai}
  \country{China}
}
\email{qysaltyfish@sjtu.edu.cn}

\author{Yiming Liu}
\orcid{0009-0009-3380-392X}
\affiliation{%
  \institution{Shanghai Jiao Tong University}
  \city{Shanghai}
  \country{China}
}
\affiliation{%
  \institution{Shanghai Innovation Institute}
  \city{Shanghai}
  \country{China}
}
\email{liu\_yiming@sjtu.edu.cn}

\author{Jiaxin Chang}
\orcid{0009-0001-0199-1059}
\affiliation{%
  \institution{Shanghai Jiao Tong University}
  \city{Shanghai}
  \country{China}
}
\email{cjx001234@sjtu.edu.cn}

\author{Weinan Zhang}
\orcid{0000-0002-0127-2425}
\affiliation{%
  \institution{Shanghai Jiao Tong University}
  \city{Shanghai}
  \country{China}
}
\affiliation{%
  \institution{Shanghai Innovation Institute}
  \city{Shanghai}
  \country{China}
}
\email{wnzhang@sjtu.edu.cn}

\author{Linpeng Huang}
\orcid{0000-0002-1531-7962}
\affiliation{%
  \institution{Shanghai Jiao Tong University}
  \city{Shanghai}
  \country{China}
}
\email{lphuang@sjtu.edu.cn}

\renewcommand{\shortauthors}{J. Liu, Y. Lin, C. Liu, Y. Qian, Y. Liu, J. Chang, W. Zhang, and L. Huang}

\begin{abstract}
Issue localization, which identifies code locations requiring modification from issue descriptions, is a critical step in automated software maintenance. Existing approaches predominantly attempt to directly align issue descriptions with code elements, yet often struggle due to the inherent abstraction gap between the issue description and code implementation.
Seeking alternative signals, our theoretical analysis suggests that test suites can serve as executable proxies for requirements, reducing localization uncertainty by 7.73 bits of entropy on average. A large-scale empirical study on 18 repositories validates this premise: existing tests cover 96.98\% of ground-truth files, and the two-hop pathway yields stronger semantic connectivity than direct matching in 82.4\% of cases.
Despite their potential, leveraging tests for localization faces two key challenges: the semantic gap separating issue descriptions from test identifiers, and the substantial noise in execution traces from infrastructure code. To address these, we propose \tool, which bridges the semantic gap through domain-knowledge-enhanced test representations and filters noise via hierarchical trace analysis. Experiments on SWE-bench Lite show that \tool achieves state-of-the-art performance, improving function-level Recall@1 by 41.57\% over the strongest baseline. When integrated into the Agentless pipeline, \tool resolves 17.72\% more issues, demonstrating practical downstream benefits.
\end{abstract}




\maketitle

\section{Introduction}
\label{sec:intro}
Automating issue resolution is a longstanding goal in software maintenance, and recent LLM-based tools have renewed interest in practical automation~\citep{hou2024large,github2025copilot}. 
A central step is \textbf{issue localization}, which occupies approximately two-thirds of the total debugging time according to empirical studies~\citep{bohme2017bug}.
Given an issue description and a repository (with its accompanying test suite), this task aims to identify the code locations that should be modified to implement the intended behavior.


Recent solutions typically follow a \emph{direct issue--code alignment} paradigm~\citep{chen2025locagent,ouyang2024repograph,liu2025codexgraph}. 
In practice, they either 
(i)~treat localization as a retrieval task and rank code elements by similarity to the issue description~\citep{fehr2025coret,chakraborty2025blaze}, or 
(ii)~use a structured search procedure, often powered by an LLM agent, to traverse the repository hierarchy, invoke tools, and iteratively narrow down candidate files and functions~\citep{xia2024agentless,li2025patchpilot,yang2024swe}. 
Despite their progress, these approaches still hinge on matching requirement-level language in issues to implementation-oriented identifiers in code, 
making them brittle when the issue describes behavior while the relevant code is organized by technical concerns~\citep{dit2013feature,eisenbarth2003locating}.

\begin{figure}[t]
    \centering
    \includegraphics[width=0.85\columnwidth]{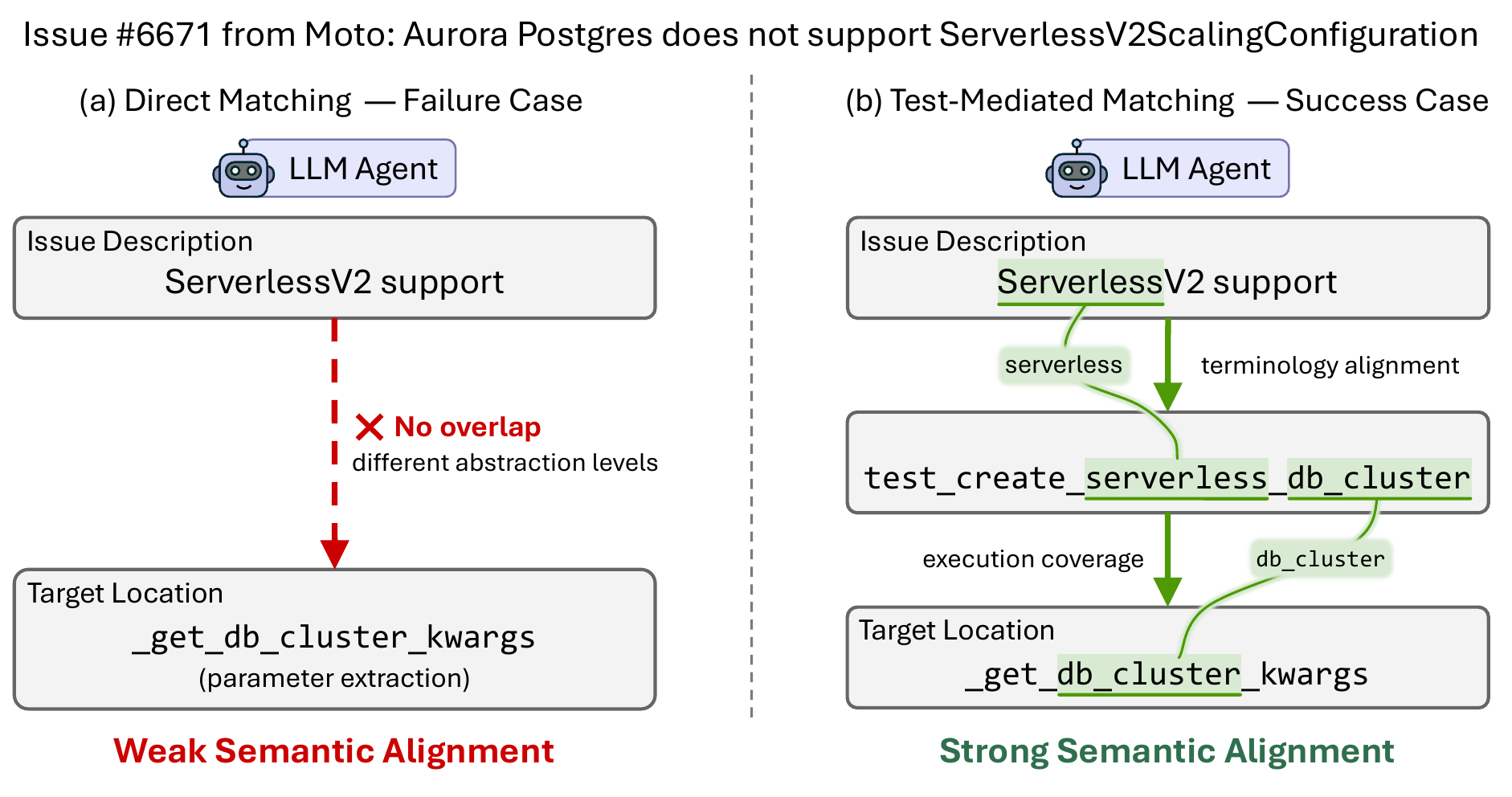}
    \caption{Motivation for test-mediated localization. (a) A failure case of direct issue-to-code matching due to weak semantic alignment. (b) A success case using tests as functional bridges to align requirements with target locations.}
    \label{fig:introduction}
    \vspace{-5pt}
\end{figure}

Concretely, issues express \textit{behavioral expectations} while code identifiers reflect \textit{technical organization}. 
As illustrated in \autoref{fig:introduction}(a), the issue requests support for ``ServerlessV2'' (a functional capability), 
yet the target function \texttt{\_get\_db\_cluster\_kwargs} is named after its implementation role in parameter extraction, 
with no indication of the feature it serves. 
A natural alternative would be requirement-aware localization, 
but explicit requirement documentation drifts from implementation and rarely maps to code precisely~\citep{levin2017co}. 
Tests, however, must remain synchronized with implementation to pass~\citep{zaidman2011studying}. 
As \autoref{fig:introduction}(b) shows, the test \texttt{test\_create\_serverless\_db\_cluster} shares behavioral terminology with the issue (both reference ``serverless'' and ``db\_cluster''), 
providing a natural intermediate target for retrieval. Executing the retrieved tests further yields execution traces that connect the test intent to the exercised implementation, including the target function. 
This two-hop pathway (issue $\rightarrow$ tests $\rightarrow$ code) can provide a more reliable evidence chain than direct issue--code matching alone. 
This is not an isolated case: our theoretical analysis suggests that test-driven localization reduces uncertainty by 7.73 bits on average compared to direct retrieval (\autoref{sec:theory}). To validate this premise, our empirical study on 18 high-quality open-source Python repositories (\autoref{sec:empirical}) shows that existing tests cover 96.98\% of ground-truth files and 66.70\% of functions, while the two-hop pathway yields stronger semantic connectivity than direct matching in 82.4\% of cases.


This effectiveness stems from tests' unique position in software development:
they are both human-readable specifications and machine-executable validators~\citep{beck2003test,hutton2009clean,adzic2011specification}.
Each test verifies specific functionality, with identifiers that encode requirement-level semantics~\citep{nagappan2008realizing,pinto2012understanding},
while execution traces establish deterministic links to implementation~\citep{ivankovic2019code,white2020establishing,eisenbarth2003locating}. 
This duality motivates our key insight: \textbf{test suites can serve as executable requirements}, as test names and assertions provide a requirement-level semantic channel, while execution traces provide a concrete dynamic link from tests to the exercised code.

Leveraging test coverage for issue localization introduces two technical challenges: 
\begin{itemize}[leftmargin=*]
    \item \textbf{C1. Domain-Specific Semantic Gap:} Retrieving relevant tests from issues requires domain knowledge absent from 
    generic embeddings, e.g., retrieving \texttt{test\_tz\_aware\_datetime} from ``connecting wrong time displayed for users in different countries'' requires recognizing that ``tz'' denotes timezone~\citep{niu2025deep,xiao2019improving,ciborowska2022fast}; 

    \item \textbf{C2. Infrastructure Noise and Trace Dilution:} 
    Localizing target code from test coverage is non-trivial, as a single test may execute thousands of functions, including infrastructure code irrelevant to the requirement. Moreover, existing tests are all passing before issue resolution, leaving no failing-test information required by spectrum-based fault localization techniques~\citep{jones2005empirical,abreu2006evaluation}. 
\end{itemize}


To address these challenges, we propose \tool (Issue Executable), 
a test-mediated localization framework that treats test suites as executable realizations of issue requirements. 
In particular, \tool tackles \textbf{C1} by retrieving candidate tests that 
reflect the issue intent using test representations enhanced with 
project-specific domain knowledge (e.g., abbreviations and API aliases) 
mined from historical commits. 
It then tackles \textbf{C2} by leveraging the runtime execution traces 
of the retrieved tests and modeling their execution hierarchy as a trace graph, 
which helps filter incidental infrastructure and highlight requirement-central code locations. 
\tool combines
(1)~domain knowledge enhancement via historical commit mining for robust issue--test alignment; and
(2)~dynamic trace graph modeling for hierarchy-aware trace denoising and localization.

Together, these components construct a focused, requirement-centric search space that maintains high recall while substantially reducing context size. On SWE-bench Lite~\citep{jimenez2023swe}, \tool boosts the Recall@1 localization performance at the file, module, and function levels by 17.78\%, 25.98\%, and 41.57\%, respectively.
When integrated into the Agentless pipeline~\citep{xia2024agentless}, our approach 
resolves 17.72\% more issues, indicating benefits for downstream patch generation.

Our contributions are as follows:
\begin{itemize}[leftmargin=*, parsep=0pt, topsep=0pt]
    \item We formalize and operationalize the notion of \textbf{tests as executable requirements} for issue localization, showing how semantic alignment at the test level and execution grounding at runtime jointly bridge the issue–code gap.
    
    \item We propose \tool, a procedure-based framework that enhances test representations with domain knowledge for effective issue-test alignment and leverages execution trace hierarchy to pinpoint suspicious locations.
    
    \item We demonstrate state-of-the-art issue localization performance on SWE-bench Lite, with 17.72\% improvement in end-to-end resolution when integrated into Agentless.
\end{itemize}

The source code, prompts, and additional materials are available on the artifact page~\cite{homepage}.

\section{Theoretical Motivation}

\label{sec:theory}

We conduct a preliminary theoretical analysis and empirical validation to establish whether the test-driven localization paradigm yields meaningful uncertainty reduction over direct issue-to-code matching. We model how uncertainty evolves along the hierarchical retrieval process (issue $\rightarrow$ tests $\rightarrow$ trace-guided code) and validate the predicted gain under method-agnostic conditions~\citep{feyzi2019inforence}. Our goal is to quantify the uncertainty reduction achievable by the test-driven paradigm itself, independent of specific implementation choices, thereby providing principled justification for the subsequent framework design.

\subsection{Formal Preliminaries}
\noindent\textbf{Setup and Notation.}
Given an issue description $d$ and repository $R$, let $L$ be the universe of candidate locations (e.g., all functions/methods), with $N = |L|$. 
Let $\mathcal{G}\subseteq L$ denote the ground-truth edit set for $d$, with $|\mathcal{G}|>0$.
For any issue localization system that returns a ranked list of locations under a predefined budget $k_{\text{ret}}$;
the resulting set is denoted as $L_{\text{ret}}$, where $|L_{\text{ret}}| = k_{\text{ret}}$.

For the test-driven pipeline, we retrieve a fixed-size set of tests $\mathcal{T}_d$, based on the given issue description $d$, with a predefined budget $k_{\text{test}}$, i.e., $|\mathcal{T}_d| = k_{\text{test}}$.
Each retrieved test $t \in {\mathcal{T}}_d$
has a dynamic coverage set $\text{Cov}(t) \subseteq L$.
The trace-constrained candidate subspace is the union coverage:
\begin{equation}
L_{\text{cov}} = \bigcup_{t\in {\mathcal{T}}_d}\text{Cov}(t),
\qquad N_{\text{cov}} = |L_{\text{cov}}|.
\label{eq:subspace_union}
\end{equation}

\noindent\textbf{Entropy Proxy.}
We approximate uncertainty with Hartley entropy $H(\mathcal{C}) \approx \log_2 |\mathcal{C}|$~\citep{hartley1928transmission, cover1999elements}, 
where $\mathcal{C}$ denotes a candidate set and $|\mathcal{C}|$ represents the size of the search space. 
In our issue localization scenario, $\mathcal{C}$ corresponds to the candidate code locations under consideration.
Our objective is not to estimate the intractable distribution $P(L\mid d)$, but to quantify the \emph{relative} uncertainty reduction induced by hierarchical retrieval and trace constraints.

\subsection{Quantifying Localization Uncertainty}

We now formulate the localization uncertainty for two different paradigms to quantify the theoretical advantage of test-driven localization. 
Specifically, we define $H_{\text{direct}}$ for the baseline approach that
localizes code locations by directly matching the issue description to code,
and $H_{\text{indirect}}$ for our proposed two-stage hierarchical process that
leverages tests as executable requirements to perform indirect localization via
an issue-to-test and test-to-code mapping.

\noindent\textbf{Direct Retrieval Uncertainty.}
A direct method ranks the entire space $L$ and returns $L_{\text{ret}}$.
Let $g_{\text{ret}} = |\mathcal{G}\cap L_{\text{ret}}|$ be the number of ground-truth locations already included in the returned set.
We define a partition-based entropy proxy that accounts for whether ground-truth locations fall inside or outside the returned set:

\begin{equation}
\begin{aligned}
H_{\text{direct}} ={}&
\mathbb{1}[g_{\text{ret}} > 0]\cdot \Big(-\log_2 \frac{g_{\text{ret}}}{|L_{\text{ret}}|}\Big) +
\mathbb{1}[|\mathcal{G}|-g_{\text{ret}} > 0]\cdot
\Big(-\log_2 \frac{|\mathcal{G}|-g_{\text{ret}}}{N-|L_{\text{ret}}|}\Big)
\end{aligned}
\label{eq:direct_entropy}
\end{equation}

Intuitively, the first term measures the ambiguity among returned candidates that contain (some of) the true edits, while the second term captures the residual uncertainty when some ground-truth edits are not retrieved.

\noindent\textbf{Test-Driven Retrieval Uncertainty.}
Test-driven localization decomposes the search into two stages: (i) retrieving requirement-relevant tests, and (ii) localizing within a trace-constrained subspace. 
We model uncertainty as the sum of stage-wise entropies, accounting for realistic failure modes.

\textbf{Stage 1: Test selection entropy.}
Among the retrieved tests $\mathcal{T}_d$, we call a test \emph{effective} if it covers at least one ground-truth location, i.e., $\text{Cov}(t)\cap \mathcal{G}\neq\emptyset$.
Let $n$ be the number of effective tests in ${\mathcal{T}}_d$.
We define:
\begin{equation}
H_{\text{stage1}} = \mathbb{1}[n>0]\cdot \Big(-\log_2 \frac{n}{k_{\text{test}}}\Big).
\label{eq:stage1_entropy}
\end{equation}
\begin{figure}

    \centering
    \includegraphics[width=0.50\linewidth]{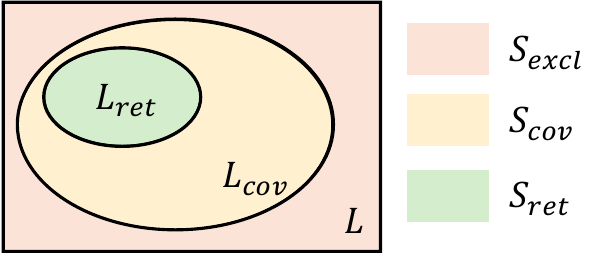}
    \caption{Venn diagram illustrating the hierarchical search space partition}
    \label{fig:venn}
    \vspace{-5pt}
\end{figure}
This term captures how confidently the pipeline selects tests that are functionally linked to the true edits. When $n=0$, the pipeline fails to retrieve any effective test, and the second stage will necessarily incur a large penalty (defined below).

\textbf{Stage 2: Trace-constrained localization entropy.}
We partition the global space $L$ into three disjoint regions induced by the trace subspace and the returned top-$k_{\text{ret}}$ locations. Critically, in the test-driven pipeline, the final retrieved set $L_{\text{ret}}$ is a subset of the trace-covered space, i.e., $L_{\text{ret}} \subseteq L_{\text{cov}}$. We define the partitions as:
\begin{equation}
S_{\text{ret}} = L_{\text{ret}}, 
\qquad S_{\text{cov}} = L_{\text{cov}} \setminus L_{\text{ret}},
\qquad S_{\text{excl}} = L\setminus L_{\text{cov}}.
\end{equation}
Here, $S_{\text{excl}}$ represents the excluded regions that are not covered by any retrieved test traces, as shown in \autoref{fig:venn}.
Let $N_i = |S_i|$ and $g_i = |\mathcal{G}\cap S_i|$ for $i\in\{\text{ret},\text{cov},\text{excl}\}$.
We then define:
\begin{equation}
H_{\text{stage2}} =
\sum_{i\in\{\text{ret},\text{cov},\text{excl}\},\, g_i>0}
\Big(-\log_2 \frac{g_i}{N_i}\Big).
\label{eq:stage2_entropy}
\end{equation}
This formulation makes two realistic phenomena explicit:
(i)~\emph{Space compression}~\citep{smytzek2025execution}: when $L_{\text{cov}}$ is much smaller than $L$, uncertainty shrinks because $N_{\text{cov}} \ll N$;
(ii)~\emph{Coverage/retrieval penalty:} if $g_{\text{excl}}>0$ (i.e., some true edits fall outside trace coverage), the term $-\log_2(g_{\text{excl}}/N_{\text{excl}})$ can dominate, reflecting the high uncertainty caused by incomplete trace constraints.
We refer to this condition as \emph{external failure} hereafter.

Finally, the hierarchical uncertainty is:
\begin{equation}
H_{\text{indirect}} = H_{\text{stage1}} + H_{\text{stage2}}.
\label{eq:indirect_entropy}
\end{equation}

\subsection{Entropy Gain Analysis}
For each issue, we define the empirical uncertainty reduction:
\begin{equation}
\Delta H = H_{\text{direct}} - H_{\text{indirect}}.
\label{eq:entropy_gain}
\end{equation}
A positive $\Delta H$ indicates that introducing tests and traces reduces the ambiguity of edit localization compared to direct matching, \emph{under the same budget} of returning top-$k_{\text{ret}}$ locations.
The decomposition further reveals two complementary directions for maximizing entropy gain:
(1)~improving test retrieval precision to increase $n/k_{\text{test}}$ and reduce the external failure rate ($g_{\text{excl}}>0$), thereby lowering $H_{\text{stage1}}$;
(2)~refining trace analysis to shrink $N_{\text{cov}}$ while preserving $g_{\text{ret}}$, thereby lowering $H_{\text{stage2}}$.
These insights directly inform our framework design in subsequent sections.

\subsection{Theoretical Validation}
We compute the above entropy proxies on 340 real-world resolved issues.
To ensure a fair comparison that isolates the effect of the paradigm itself, we employ a unified embedding model for both retrieval pathways: (i)~\emph{direct retrieval}, which ranks code locations in $L$ by similarity to the issue description, and (ii)~\emph{indirect retrieval}, which first retrieves tests by the same similarity measure, then constrains the candidate space via execution coverage.
Test coverage is obtained from standard execution traces without task-specific filtering. 

\begin{wraptable}[16]{r}{0.5\textwidth}
\centering
\caption{Pipeline-aware entropy analysis under realistic test retrieval. 
$\Delta H$ is the uncertainty reduction (\autoref{eq:entropy_gain}). 
``External failure'' reports the fraction of issues with $g_{\text{excl}}>0$ (some ground-truth edits outside trace coverage). Results use $k_{\text{test}}=5$ and $k_{\text{ret}}=10$.}
\label{tab:entropy_bounds}

\begin{small}
\begin{tabular}{lrrrr}
    \toprule
    \textbf{Metric} & \textbf{Value} \\
    \midrule
    Total Valid Samples & 340 \\
    Mean $H_{\text{direct}}$ (bits) & 11.90 \\
    Mean $H_{\text{indirect}}$ (bits) & 4.17 \\
    Mean $H_{\text{stage1}}$ (bits) & 0.14 \\
    Mean $H_{\text{stage2}}$ (bits) & 4.04 \\
    \midrule
    \textbf{Mean Entropy Gain $\Delta H$ (bits)} & \textbf{7.73} \\
    External failure rate ($g_{\text{excl}}>0$) & 24.41\% \\
    \bottomrule
\end{tabular}
\end{small}
\vspace{-5pt}
\end{wraptable}

The validation results are shown in \autoref{tab:entropy_bounds}.
A consistently positive $\Delta H$ confirms that the two-hop pathway (issue $\rightarrow$ tests $\rightarrow$ trace-constrained code) reduces localization uncertainty on average.
Notably, even with a high external failure rate (24.41\%), the overall $\Delta H$ remains strongly positive.
For issues without external failure, trace constraints yield substantial gains ($\Delta H = 9.21$ bits) by compressing the search space.
For issues with external failure, direct retrieval often fails to assign high priority to the ground-truth locations ($H_{\text{direct}}=11.90$ bits),
while indirect retrieval still achieves partial hits within the trace-constrained subspace, resulting in a positive $\Delta H$ of 3.15 bits.
These findings validate the theoretical advantage of test-driven localization and motivate the design of \tool, which realizes this paradigm through enhanced test retrieval and hierarchical trace analysis.

\section{Empirical Study}
\label{sec:empirical}

Theoretical arguments in \autoref{sec:theory} suggest that tests can serve as \emph{executable requirements} for issue localization by providing (i) requirement-level semantics and (ii) a dynamic link to exercised code.
In this section, we validate this premise empirically on a large-scale dataset, and quantify when and how tests provide actionable signals for localization.
We test the following hypotheses:
\begin{itemize}[leftmargin=*, itemsep=2pt]
    \item \textbf{H1 (Coverage feasibility).} Existing tests execute the ground-truth edit locations.
    \item \textbf{H2 (Retrievability).} Tests that cover ground-truth locations are more semantically aligned with the issue than non-covering tests.
    \item \textbf{H3 (Bridging effect).} The two-hop pathway Issue$\rightarrow$Test$\rightarrow$Location provides stronger semantic connectivity than direct Issue$\rightarrow$Location matching.
\end{itemize}

\subsection{Setup}
\label{sec:empirical_setup}

The empirical investigation is conducted on a collection of 929 issues sourced from SWE-bench~\citep{jimenez2023swe} and SWE-bench Gym~\citep{pan2024training}, covering 18 widely-adopted Python repositories (each with $\geq$500 stars). These projects represent diverse domains including web frameworks, data processing, and scientific computing. To ensure the reliability of dynamic analysis, we employ a Docker-based execution framework. Each repository-issue pair is isolated within a container where the environment is reset to its pre-patch state. We retain only those instances where the test environment is sufficiently robust, defined as having at least 50\% of test functions execute successfully, to prevent environment configuration errors from confounding the findings.

Function-level coverage traces are collected via instrumentation using the Python \texttt{sys.settrace} mechanism. During the execution of the full test suite, the tracer records call and return events for every function invocation. We apply a filtering layer to exclude standard library calls and third-party dependencies, retaining only repository-internal functions identified by their fully-qualified paths. Ground-truth edit locations are extracted by parsing the pull request diffs associated with each issue, identifying the specific functions modified by developers to resolve the reported problem.

For semantic analysis, the \texttt{bge-large-en-v1.5}~\citep{bge_embedding} model is utilized to generate dense vector embeddings. Each test case is transformed into a textual representation by concatenating its docstring, its fully-qualified module path, and its complete function body. These components are truncated to 512 tokens to fit the model context window. Code locations are similarly represented by their fully-qualified signatures. Semantic similarity is then quantified using the cosine similarity between the resulting embeddings.

\subsection{Findings}
\label{sec:empirical_findings}

\begin{figure*}[t]
    \centering
    \begin{minipage}{0.48\textwidth}
        \centering
        \includegraphics[width=0.9\textwidth]{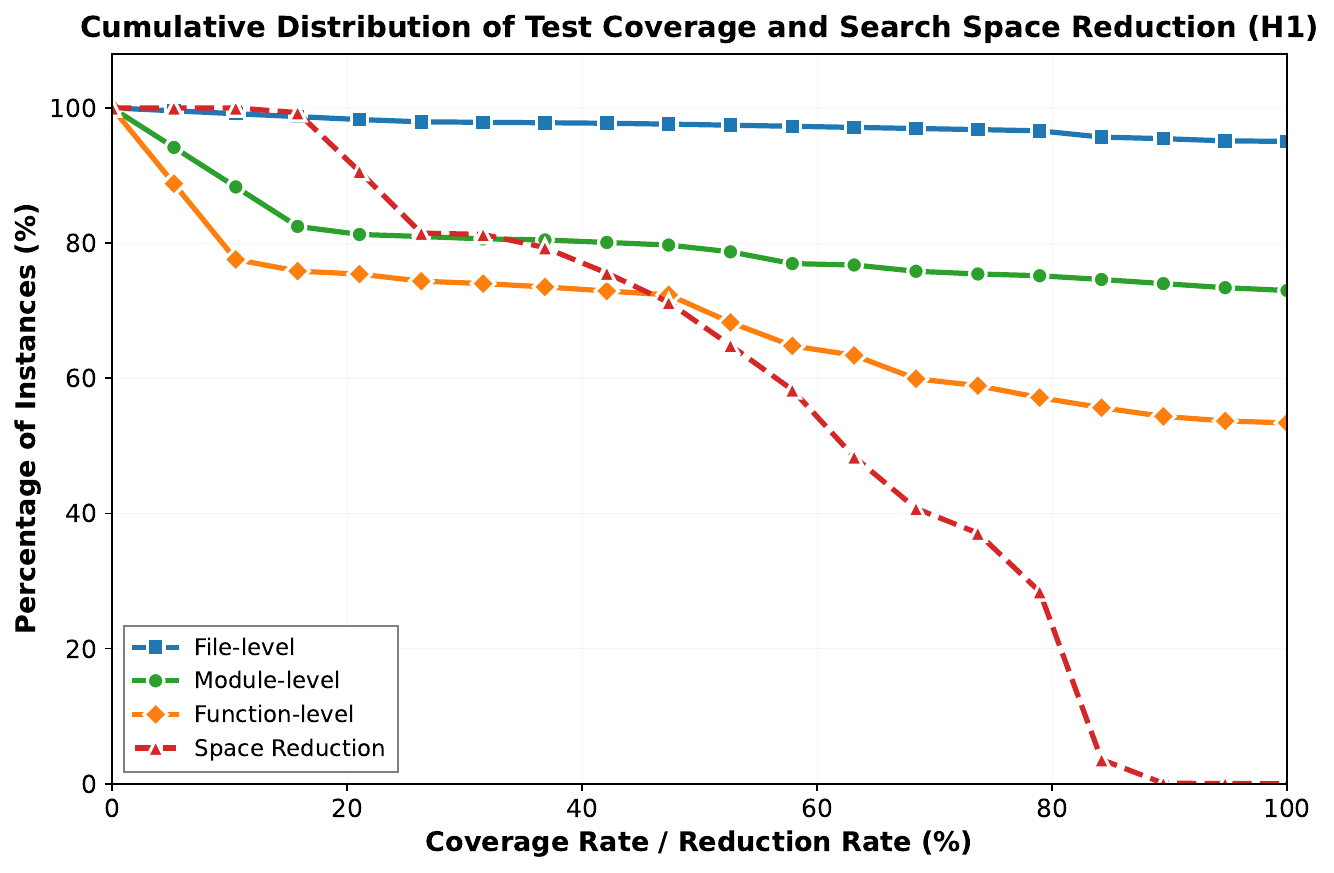}
        \caption{Cumulative distribution of test coverage rates (H1). Existing tests cover 96.98\% of ground-truth files and 66.70\% of ground-truth functions on average,  reducing search space to 58.39\% of the repository.}
        \label{fig:empirical_h1}
    \end{minipage}
    \hfill
    \begin{minipage}{0.48\textwidth}
        \centering
        \includegraphics[width=\textwidth]{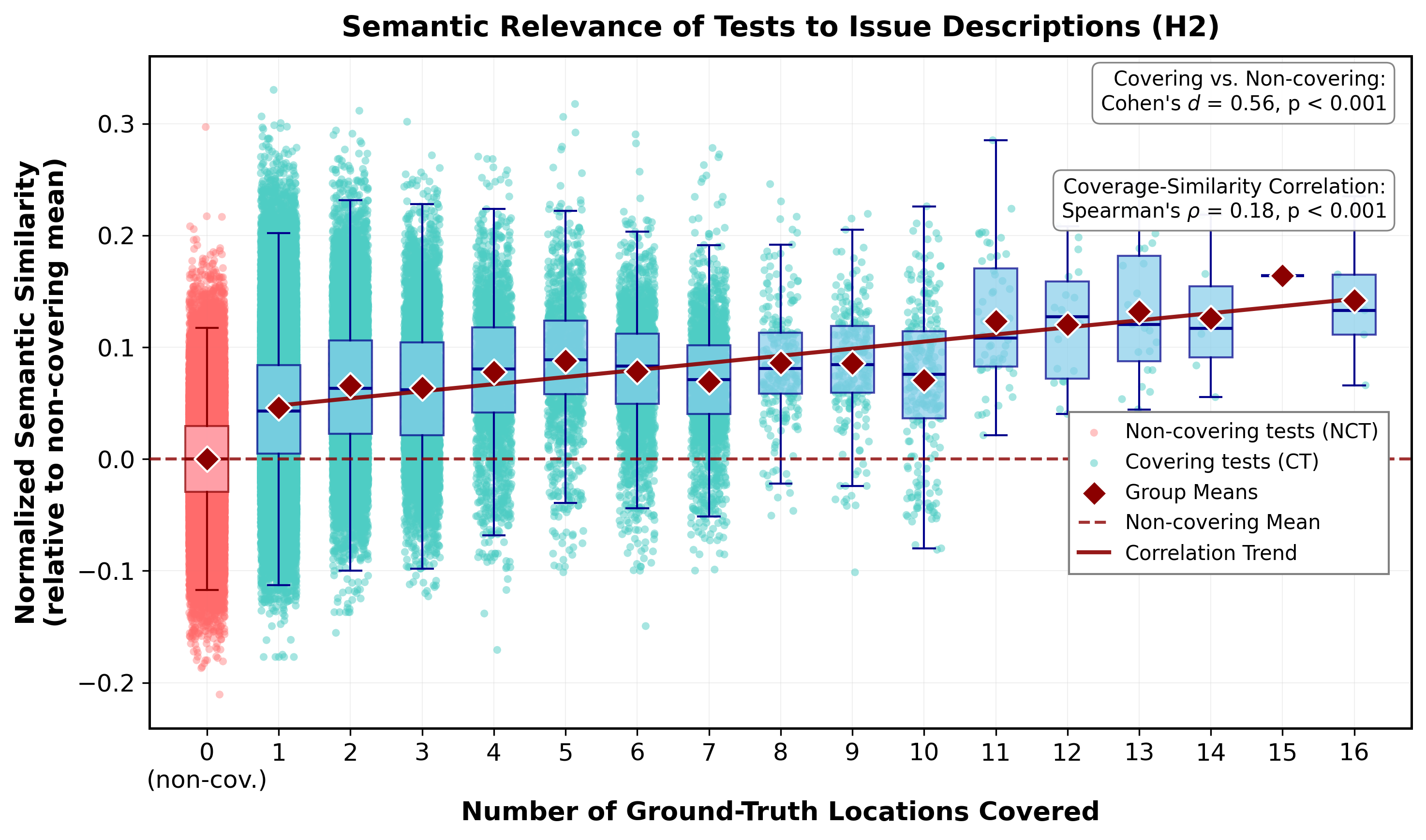}
        \caption{Semantic similarity of tests to issue descriptions (H2). Tests covering more ground-truth locations exhibit higher similarity to the issue, enabling retrieval-based selection.}
        \label{fig:empirical_h2}
    \end{minipage}
    \vspace{-5pt}
\end{figure*}

\textbf{H1. (Coverage feasibility).}
\autoref{fig:empirical_h1} demonstrates the structural feasibility of using existing tests to reach target locations. By intersecting the dynamic traces of the entire test suite with the ground-truth edit sets, we observe that existing tests provide broad coverage, reaching 96.98\% of files and 66.70\% of functions requiring modification. This confirms that for the vast majority of issues, the necessary execution signals already exist within the repository infrastructure. Furthermore, the search space reduction is quantified by comparing the cardinality of the union of all test-covered functions against the total functions in the repository. This restriction narrows the candidate pool to 58.39\% of the repository on average, achieving high recall while providing effective search space compression for subsequent localization stages.

\noindent\textbf{H2. (Retrievability).}
For test-based localization to succeed, covering tests must be distinguishable from non-covering tests via semantic signals. Our experimental setup compares the similarity of tests that cover ground-truth locations against a baseline of randomly sampled tests from the same repository. After verifying the normality of the similarity distribution via the Shapiro-Wilk test, a paired t-test is conducted~\citep{shapiro1965analysis}. As shown in \autoref{fig:empirical_h2}, covering tests exhibit significantly higher similarity to issue descriptions in 90.9\% of instances ($p < 0.001$; Cohen's $d = 0.56$), indicating a medium-to-large statistical effect. Additionally, Spearman's rank correlation analysis between the number of ground-truth locations covered and the issue similarity yields $\rho = 0.18$ ($p < 0.001$). This suggests that semantic relevance can effectively prioritize tests with broader functional coverage of the requirement.

\noindent\textbf{H3. (Bridging effect).}
Our central hypothesis is that tests serve as semantic bridges between abstract issues and technical code. For each (issue, location) pair where coverage exists, we compare the direct similarity $s_{\text{direct}} = \text{sim}(\text{issue}, \text{location})$ against a test-mediated pathway strength. We formalize this mediated connectivity using a two-hop geometric mean formulation~\citep{feyzi2019inforence}:
$$s_{\text{mediated}} = \max_{t \in \mathcal{T}_{\text{cover}}} \sqrt{\text{sim}(\text{issue}, t) \cdot \text{sim}(t, \text{location})}$$
This formulation ensures that a strong bridge requires balanced semantic associations across both the requirement-to-test and test-to-implementation links. As illustrated in the results, the mediated pathway is stronger in 82.4\% of cases (paired $t$-test, $p < 0.001$; Cohen's $d = 0.93$). The substantial effect size confirms that tests act as effective semantic relays, leveraging both requirement-level language and deterministic execution grounding to achieve connectivity that direct issue-to-code matching cannot.

\subsection{Implications for \tool}
\label{sec:empirical_implications}

These findings directly inform \tool's design.
H1 establishes that test coverage provides a high-recall, reduced-entropy search space.
H2 confirms that semantic retrieval can identify\linebreak requirement-relevant tests, though the moderate effect size ($d = 0.56$) motivates our domain-knowledge enhancement (\autoref{sec:domain}) to strengthen Issue$\rightarrow$Test alignment.
H3 validates the two-hop localization strategy, while the gap between covering and non-covering code within traces motivates our hierarchical trace analysis (\autoref{sec:trace}) to filter infrastructure noise.\par

\section{The \tool Framework}
\label{sec:method}

The previous two sections provide complementary support for our central premise that \emph{tests can serve as executable requirements} for issue localization: \autoref{sec:theory} motivates the premise from a theoretical perspective, and \autoref{sec:empirical} validates it empirically at scale.
Building on this premise, we introduce \tool, which operationalizes tests as an intermediate layer that connects issue descriptions to candidate code locations.

\subsection{Problem Formulation}
\label{sec:formulation}

Given a natural-language issue description $d$ and a repository $R$ with code entities $V$, issue localization aims to identify the code locations $L = \{l_1, l_2, \ldots, l_k\} \subseteq V$ that need to be modified to resolve the issue~\citep{wong2016survey,niu2025deep}. Formally, we seek the optimal $L^*$ that minimizes $|L|$ such that $\text{Patch}(R, L) \models d$.

Guided by the preliminary study in \autoref{sec:theory}, which confirms the uncertainty reduction achievable via test-driven localization, we instantiate a two-stage formulation. First, we retrieve tests $\mathcal{T}_d$ semantically aligned with $d$ to maximize the effective test ratio $n/k_{\text{test}}$. Second, we leverage the execution coverage $\text{Cov}(\mathcal{T}_d)$ to localize $L^*$ within a trace-constrained, low-entropy search space.

\begin{figure*}
    \centering
    \includegraphics[width=\textwidth]{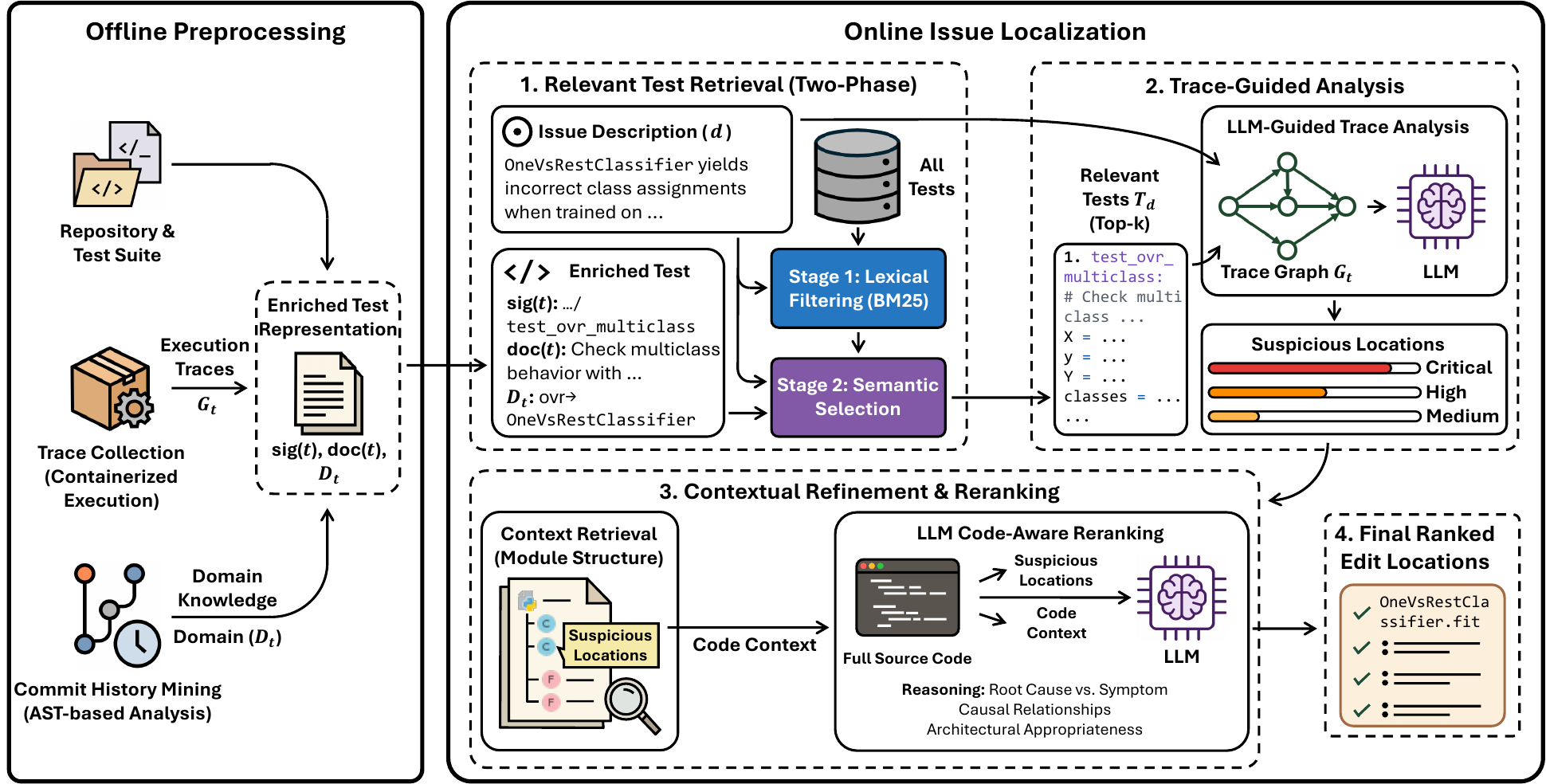}
    \caption{Overview of the \tool framework. \textbf{Left:} Offline preprocessing collects test execution traces $G_t$ via containerized execution and extracts domain knowledge $D_t$ from commit history to construct enriched test representations. \textbf{Right:} Online issue localization proceeds through 3 steps: (1)~relevant test retrieval combining BM25 lexical filtering and LLM-based semantic selection to obtain $\mathcal{T}_d$, (2)~trace-guided analysis leveraging execution traces and LLM reasoning to identify suspicious locations with confidence levels, (3)~contextual refinement and reranking using module structure and full source code, and outputs final ranked edit locations $L^*$ for downstream patch generation.}
    \label{fig:framework}
\end{figure*}

\noindent\textbf{Framework Overview.}
\autoref{fig:framework} illustrates the framework. \tool first performs offline preprocessing to collect test execution traces and mine commit history for enriching test representations with domain knowledge (\autoref{sec:domain}). At inference time, given an issue description, the localization pipeline proceeds as following stages: \ding{182}~retrieve relevant tests $\mathcal{T}_d$ through two-phase filtering (\autoref{sec:retrieval}), implementing the first stage of our formulation; \ding{183}~analyze execution traces to identify suspicious locations within $\text{Cov}(\mathcal{T}_d)$ (\autoref{sec:trace}); \ding{184}~refine and rerank candidates to produce the final $L^*$ (\autoref{sec:refine}).

\subsection{Domain Knowledge Enhancement}
\label{sec:domain}

The semantic distance between the generic embeddings of the issue description and the tests remains substantial, 
as default test representations (e.g., function signatures and docstrings) fail to capture domain-specific associations and project-specific terminology.
By incorporating domain knowledge $D_t$, we bridge the terminology gap between issue descriptions and test identifiers, effectively reducing the semantic gap for requirement-relevant tests.

To this end, we extract domain knowledge from the repository's commit history, which captures the semantic context under which tests were introduced and evolved. For each test function $t$, we extract two types of signals.

First, we identify the commit that first introduced $t$ via AST-based diff analysis, filtering out low-quality commits (e.g., merge commits, bulk refactoring). The associated commit message provides requirement-level semantic context:
\begin{equation}
m_t = \text{msg}(\argmin_{h \in \mathcal{H}} \{ \text{time}(h) \mid t \in \text{added}(h) \})
\label{eq:commit_msg}
\end{equation}
where $\mathcal{H}$ denotes the commit history and $\text{added}(h)$ returns test functions introduced in commit $h$.

Second, we identify co-changed entities, i.e., code locations frequently modified together with $t$ across commits:
\begin{equation}
\mathcal{A}_t = \{ l \in V \mid \text{cochange}(t, l) \geq \tau \}
\label{eq:cochange}
\end{equation}
where $V$ is the set of all code entities, $\text{cochange}(t, l)$ counts the number of commits in which both $t$ and $l$ were modified, and $\tau$ is a frequency threshold. These entities typically represent requirement-related code that $t$ implicitly covers.


Finally, we distill domain-specific tokens from these extracted signals to form the test-specific knowledge set $D_t$:

\begin{equation}
D_t = \text{tok}(m_t) \cup \text{ident}(\mathcal{A}_t) \setminus \text{tok}(\text{sig}(t))
\label{eq:domain_tokens}
\end{equation}
where $\text{tok}(\cdot)$ extracts semantic tokens from the commit message and $\text{ident}(\mathcal{A}_t)$ extracts function and class names from co-changed entities. The set difference removes tokens already present in the test signature to avoid redundancy.

We construct the enriched representation by concatenating three components for each test:
\begin{equation}
\text{repr}(t) = [\text{sig}(t); \text{doc}(t); D_t]
\label{eq:repr}
\end{equation}
where $\text{sig}(t)$ denotes the fully qualified test identifier and $\text{doc}(t)$ is the docstring if present. 
By incorporating $D_t$, domain-specific tokens provide additional semantic alignment between issues and tests beyond what generic embeddings can capture from $\text{sig}(t)$ alone, thereby more effectively bridging the terminology gap and reducing the semantic gap for requirement-relevant tests.

\subsection{Relevant Test Retrieval}
\label{sec:retrieval}

This section implements the first stage of our localization pipeline.
Given an issue description $d$ and the enriched test representations constructed in \autoref{sec:domain}, 
\tool 
employs a two-phase filtering process that 
retrieves a subset $\mathcal{T}_d \subseteq \mathcal{T}$ with $|\mathcal{T}_d| \ll |\mathcal{T}|$ while maximizing information extraction for downstream localization from the issue.

In the lexical filtering phase, we apply BM25~\citep{robertson2009probabilistic} to efficiently reduce the search space from potentially thousands of tests to a manageable candidate set:
\begin{equation}
\mathcal{T}_{\text{cand}} = \text{Filter}(d, \mathcal{T}, N)
\label{eq:filter}
\end{equation}
where $\text{Filter}$ returns the top-$N$ tests ranked by lexical similarity to the issue $d$. This phase prioritizes recall, ensuring that requirement-relevant tests are retained.

In the semantic selection phase, we employ an LLM to carefully identify the most relevant subset from the candidates. Given the enriched representations $\text{repr}(t)$ that incorporate domain-specific tokens $D_t$, the LLM explicitly reasons about the issue's intent and the tests' purposes:
\begin{equation}
\mathcal{T}_d = \text{Select}(d, \{\text{repr}(t) \mid t \in \mathcal{T}_{\text{cand}}\}, k)
\label{eq:select}
\end{equation}
where $k \ll N$ controls the size of the final retrieved set. The LLM is prompted to analyze which components are likely responsible for the reported issue, which test modules exercise those components, and which specific tests are most likely to cover the root cause. 
This produces a small set of relevant tests $\mathcal{T}_d$ whose execution traces will guide subsequent localization. 
The LLM is guided by manually designed examples illustrating common pitfalls and selection criteria; the full prompt is provided on the artifact page~\cite{homepage}.

This two-phase design strikes a balance between efficiency and quality: BM25 filtering reduces the search space from $|\mathcal{T}|$ to $N$ candidates efficiently, while LLM selection further refines to $k \ll N$ high-quality tests, avoiding the prohibitive cost of applying LLM reasoning to all tests.

\subsection{Trace-Guided Localization}
\label{sec:trace}

The retrieved tests $\mathcal{T}_d$ narrow the search space from the entire test suite to a small, requirement-relevant subset.
However, test retrieval alone is insufficient for precise localization:
the execution traces of these tests typically cover a large portion of the codebase,
where the true edit locations constitute only a small fraction ($|L^*| \ll |\text{Cov}(\mathcal{T}_d)|$).
While a single test may execute hundreds of functions, its execution trace preserves caller-callee relationships that encode causal structure. 
We leverage this hierarchical information to distinguish requirement-central code from incidental infrastructure and utility functions.

For each retrieved test $t \in \mathcal{T}_d$, we extract its dynamic execution trace and construct a directed graph $G_t = (V_t, E_t)$ rooted at $t$. The nodes are functions covered by the test:
\begin{equation}
V_t = \text{Cov}(t)
\label{eq:trace_nodes}
\end{equation}
The edge set captures observed caller-callee relationships:
\begin{equation}
E_t = \{(u, v) \mid u \xrightarrow{t} v\}
\label{eq:trace_edges}
\end{equation}
where $u \xrightarrow{t} v$ denotes that $u$ directly calls $v$ during execution of $t$. Since $G_t$ may exceed LLM context limits, we apply BFS-based pruning to retain only shallow layers of the call hierarchy. 
To ensure trace quality, we collect complete execution paths only from tests that execute and pass successfully, discarding traces from tests that fail or raise errors to avoid noise from engineering failures. When the issue's root cause lies outside the coverage of retrieved tests, localization may fail, and we discuss such cases in \autoref{sec:failure_missing_coverage}.

We then analyze these execution traces in conjunction with the issue description $d$ and the enriched test representations $\text{repr}(t)$:
\begin{equation}
S_t = \text{Analyze}(d, \text{repr}(t), G_t)
\label{eq:analyze}
\end{equation}
where $S_t \subseteq V_t$ denotes the set of suspicious locations identified from trace $G_t$. 
Guided by step-by-step instructions, the LLM generates diagnostic reports that analyze potential assertion deviations, hypothesize root causes, trace error propagation paths through the call chain, and identify systematic risks across related functions.

Based on these analyses, each location $l \in S_t$ is assigned a confidence level $\text{conf}(l) \in \{\text{critical}, \allowbreak \text{high}, \allowbreak \text{medium}\}$. The aggregated suspicious set is:
\begin{equation}
S = \bigcup_{t \in \mathcal{T}_d} S_t
\label{eq:suspicious_set}
\end{equation}

By analyzing execution traces with diagnostic reasoning, we aim to acquire the suspicious set $S$, 
which is designed to be sparse ($|S| \ll |\text{Cov}(\mathcal{T}_d)|$) while maintaining high recall of ground-truth locations, filtering out infrastructure code that contributes little information about the issue's root cause.

\subsection{Refinement and Reranking}
\label{sec:refine}

Building on the trace-guided suspicious set $S$ from \autoref{sec:trace},
we further refine and rerank candidates by enriching each location
with additional contextual information.

The coverage-based candidate set may miss relevant locations that are structurally related but not directly covered.
To address this, for each file $f$ containing at least one suspicious location, we retrieve its structure as context. For modules exceeding context limits, we provide a compressed skeleton that preserves function signatures and class hierarchies while omitting implementation details. 
Here $S'$ denotes the expanded candidate set after contextual refinement, obtained via step-by-step guided reasoning over module structure.
\begin{equation}
S' = S \cup \text{Refine}(\{f \mid f \cap S \neq \emptyset\}, d)
\label{eq:refine}
\end{equation}

To prioritize candidates and filter false positives, we retrieve the full source code for each candidate location and employ the LLM to perform code-aware reranking, guided by a designed example showing how to filter non-fix locations and retain coupled locations requiring joint changes,
reasoning about root-cause distinctions, inter-candidate causalities, and the architectural appropriateness of each contextual refinement:
\begin{equation}
L^* = \text{Rerank}(S', \{\text{code}(l) \mid l \in S'\}, d)
\label{eq:rerank}
\end{equation}
where $\text{code}(l)$ denotes the full source code of location $l$. The output $L^*$ is the final ranked list of edit locations for downstream patch generation.

This refinement and reranking process ensures that $L^*$ includes not only directly covered locations but also structurally and semantically related entities, improving recall beyond execution coverage.
The reranking stage then prioritizes the most plausible edit locations, improving top-$k$ precision.

\section{Experiments}
\label{sec:experiments}

We evaluate \tool on SWE-bench Lite to answer the following research questions: 
\begin{itemize}[leftmargin=*]
    \item \textbf{RQ1 (Localization performance, \autoref{sec:main_results})} How effective is \tool for issue localization?
    
    \item \textbf{RQ2 (Issue resolution performance, \autoref{sec:downstream})} Can improved localization benefit downstream issue resolution?

    \item \textbf{RQ3 (Cost Analysis, \autoref{sec:efficiency})} 
    What is the cost-efficiency of \tool compared to existing baselines?

    
    \item \textbf{RQ4 (Ablation study, \autoref{sec:ablation})} 
    How does each component contribute to the overall performance of \tool?

\end{itemize}


\subsection{Experimental Setup}
\label{sec:setup}

\textbf{Benchmarks.}
We evaluate on a widely-adopted benchmark for automated issue resolution. SWE-bench Lite~\citep{jimenez2023swe} contains 300 issues sampled from 11 popular Python repositories, filtered for self-contained problems solvable without extensive codebase knowledge.

\noindent\textbf{Baselines.}
We compare against three categories of methods. \textit{Retrieval-based} approaches include BM25~\citep{robertson2009probabilistic}, mGTE~\citep{zhang2024mgte}, CodeSage~\citep{zhang2024code}, and CodeRankEmbed~\citep{suresh2024cornstack}. \textit{Procedure-based} methods include Agentless~\citep{xia2024agentless} and PatchPilot~\citep{li2025patchpilot}. \textit{Agent-based} approaches include LocAgent~\citep{chen2025locagent}, SWE-Agent~\citep{yang2024swe}, OrcaLoca~\citep{yu2025orcaloca}, OpenHands~\citep{wang2024openhands}, and MoatlessTools~\citep{antoniades2024swe}.

\noindent\textbf{Metrics.}
We evaluate localization at three granularities: file, module (class or top-level function), and function. We report Precision (Prec.) and Recall (Rec.) at cutoffs @1, @3 and @5, measuring the ability to identify ground-truth edit locations within the top-ranked predictions.
We use resolved rate as the metric for the downstream repair task, and report F1-score for the ablation study.


\noindent\textbf{Implementation Details.}
We use \texttt{GPT\-4o-2024-05-13} and \texttt{Claude\-3.5-sonnet-20241022} as the base models for all methods to ensure fair comparison. For \tool, we set the co-change frequency threshold $\tau$ to 3, the candidate test set size $|\mathcal{T}_{\text{cand}}|$ to 200, and the final selected test set size $|\mathcal{T}_d|$ to 5.

\subsection{RQ1. Localization Performance}
\label{sec:main_results}



\autoref{tab:lite} summarizes localization performance on SWE-bench Lite at file, module, and function granularities, reporting Precision and Recall at @1/@3/@5.

\begin{table*}[t]
\centering
\setlength{\tabcolsep}{1.5pt}  
\renewcommand{\arraystretch}{1.05}
\scriptsize
\caption{RQ1. Performance comparison on SWE-bench Lite. Precision and Recall at @1, @3, and @5 are reported in \%. \textbf{Bold}: best; \underline{Underline}: second best.}
\label{tab:lite}
\begin{tabular}{@{}ll cccc cccccc cccccc@{}}
\toprule
\multirow{3.5}{*}{\textbf{Type}} & \multirow{3.5}{*}{\textbf{Method}} 
& \multicolumn{4}{c}{\textbf{File (\%)}} 
& \multicolumn{6}{c}{\textbf{Module (\%)}} 
& \multicolumn{6}{c}{\textbf{Function (\%)}} \\
\cmidrule(lr){3-6} \cmidrule(lr){7-12} \cmidrule(l){13-18}
& 
& \multicolumn{2}{c}{Precision} & \multicolumn{2}{c}{Recall}
& \multicolumn{3}{c}{Precision} & \multicolumn{3}{c}{Recall}
& \multicolumn{3}{c}{Precision} & \multicolumn{3}{c}{Recall} \\
\cmidrule(lr){3-4} \cmidrule(lr){5-6} \cmidrule(lr){7-9} \cmidrule(lr){10-12} \cmidrule(l){13-15} \cmidrule(lr){16-18}
& 
& {\centering\tiny @1} & {\centering\tiny @3} & {\centering\tiny @1} & {\centering\tiny @3}
& {\centering\tiny @1} & {\centering\tiny @3} & {\centering\tiny @5} & {\centering\tiny @1} & {\centering\tiny @3} & {\centering\tiny @5}
& {\centering\tiny @1} & {\centering\tiny @3} & {\centering\tiny @5} & {\centering\tiny @1} & {\centering\tiny @3} & {\centering\tiny @5} \\
\midrule
\multirow{5}{*}{Retrieval} 
& BM25  
& 27.37 & 14.72 & 27.37 & 43.43
& 19.71 & 12.04 & 10.35 & 19.34 & 29.87 & 32.79
& 13.87 & 7.54 & 5.40   & 13.26 & 21.41 & 25.36 \\
& mGTE 
& 35.04 & 22.44 & 35.04 & 63.50
& 29.93 & 19.22 & 16.70 & 29.38 & 47.26 & 50.36
& 18.25 & 12.16 & 8.91  & 17.15 & 34.25 & 41.42 \\
& CodeSage  
& 32.85 & 19.34 & 32.85 & 51.09
& 26.28 & 15.88 & 14.43 & 25.91 & 35.95 & 38.87
& 14.23 & 7.91  & 6.20  & 12.77 & 21.44 & 27.86 \\
& CodeRankEmbed  
& 28.83 & 18.31 & 28.83 & 52.19
& 24.82 & 17.52 & 15.85 & 24.27 & 37.41 & 40.33
& 16.42 & 8.88  & 6.57  & 15.82 & 25.30 & 30.47 \\
\midrule
\multicolumn{18}{c}{\textbf{\llmicon{1.2em}{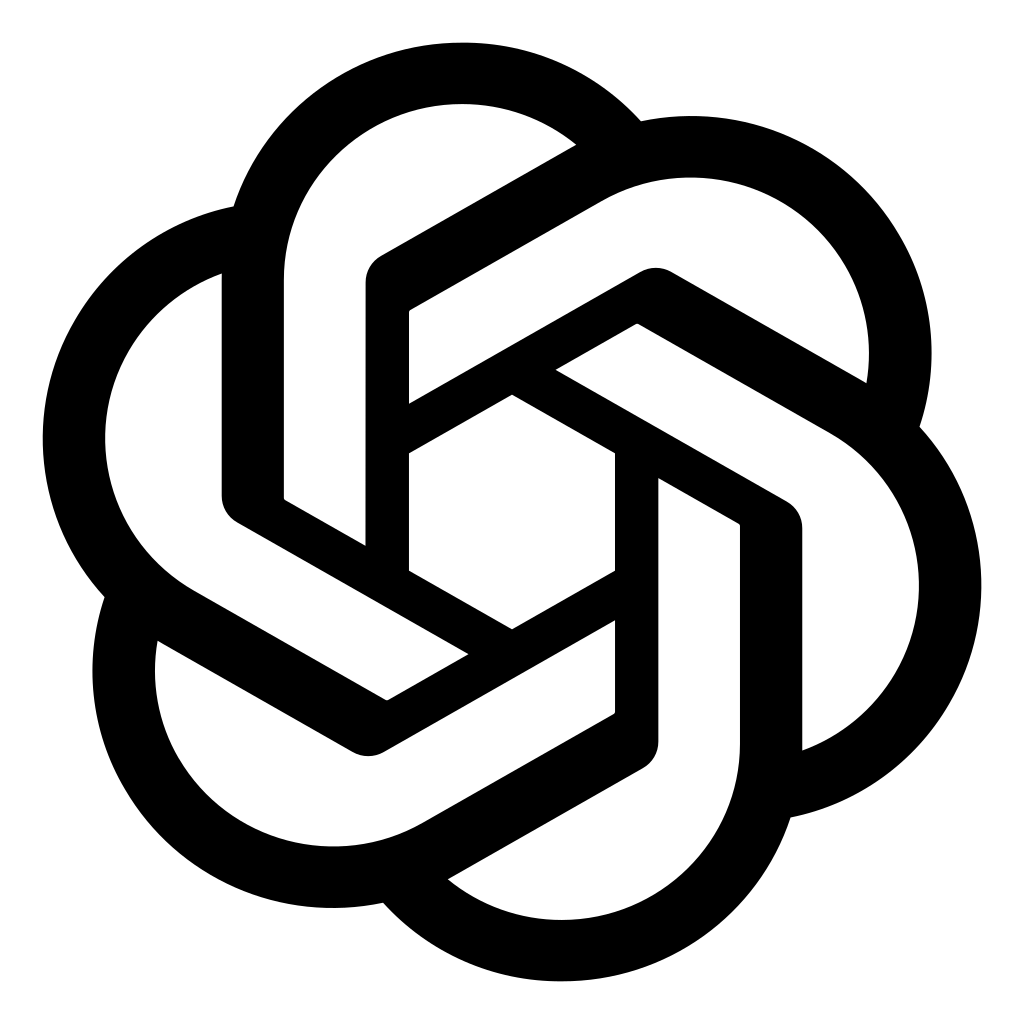}\hspace{0.35em}GPT-4o-2024-05-13}} \\
\midrule
\multirow{2}{*}{Procedure} 
& Agentless
& \underline{59.49} & 38.38 & \underline{59.49} & \textbf{78.27}
& 47.75 & 25.18 & 21.30 & \underline{46.35} & \underline{62.59} & \underline{66.42}
& 23.36 & 19.04 & 16.53 & 21.78 & \underline{43.25} & \underline{47.26} \\
& PatchPilot
& 53.70 & 39.57 & 53.70 & 64.44
& \underline{47.78} & 34.38 & 33.16 & 45.37 & 55.86 & 57.35
& \underline{30.74} & 22.78 & 21.54 & 26.98 & 34.51 & 35.25 \\
\midrule
\multirow{5}{*}{Agent} 
& LocAgent
& 57.30 & 33.21 & 57.30 & 77.74 
& 18.61 & 16.91 & 13.14 & 18.43 & 48.36 & 57.30
& 13.87 & 12.47 & 10.15 & 12.90 & 33.64 & 43.43 \\
& SWE-Agent
& 54.74 & \underline{54.01} & 54.74 & 57.66
& 41.24 & 40.82 & 40.77 & 40.15 & 43.61 & 43.61
& 30.29 & \textbf{29.50} & \textbf{29.45} & 28.25 & 32.00 & 32.00 \\
& OrcaLoca 
& 55.11 & 53.53 & 55.11 & 60.58
& 44.16 & \underline{41.91} & \underline{41.72} & 43.43 & 51.45 & 51.45
& 10.95 & 22.16 & 21.85 & 10.58 & 42.03 & 44.28 \\
& OpenHands
& 31.39 & 30.41 & 31.39 & 31.75
& 28.83 & 27.86 & 26.41 & 28.28 & 28.83 & 29.75
& 24.09 & 24.21 & 23.13 & 21.90 & 24.15 & 24.64 \\
& MoatlessTools
& 53.65 & 38.99 & 53.65 & 54.74
& 43.43 & 40.82 & 40.62 & 42.52 & 44.71 & 44.71
& 30.66 & \underline{27.74} & \underline{27.41} & \underline{29.01} & 30.60 & 31.25 \\
\midrule
Procedure
& \tool
& \textbf{70.07} & \textbf{59.43} & \textbf{70.07} & \underline{78.10}
& \textbf{60.58} & \textbf{43.98} & \textbf{43.55} & \textbf{58.39} & \textbf{68.86} & \textbf{70.32}
& \textbf{46.72} & 26.95 & 25.94 & \textbf{41.07} & \textbf{53.36} & \textbf{55.67} \\
\midrule
\multicolumn{18}{c}{\textbf{\llmicon{1.2em}{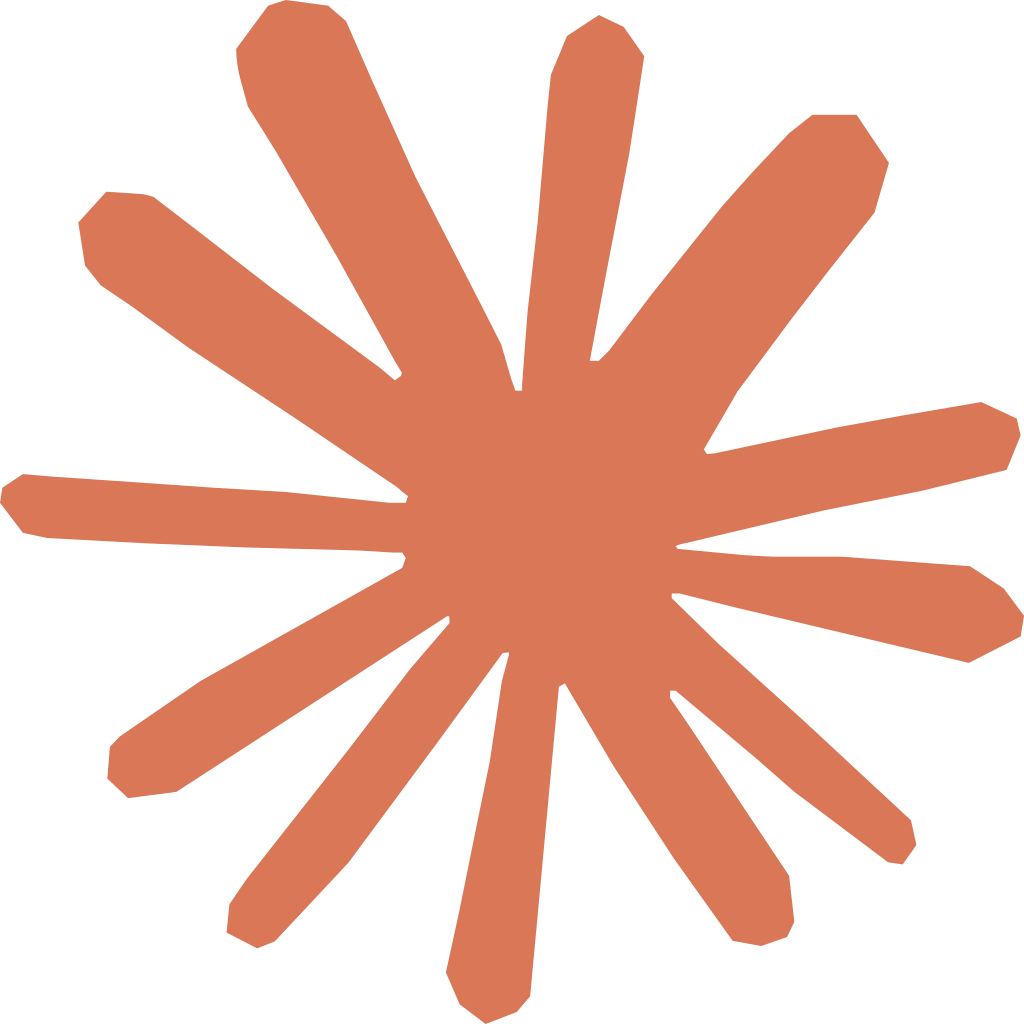}\hspace{0.35em}Claude-3-5-Sonnet-20241022}} \\
\midrule
\multirow{2}{*}{Procedure} 
& Agentless
& 57.71 & 42.57 & 57.71 & 69.70
& \underline{57.66} & 26.88 & 21.83 & \underline{55.47} & \underline{67.67} & \textbf{69.13}
& 24.45 & 17.09 & 13.87 & 21.09 & 40.03 & 43.65 \\
& PatchPilot
& 55.84 & 37.29 & 55.84 & 67.15
& 50.73 & 26.15 & 22.79 & 48.97 & 56.69 & 59.18
& 24.09 & 15.27 & 13.75 & 20.67 & 28.45 & \textbf{53.66} \\
\midrule
\multirow{5}{*}{Agent} 
& LocAgent
& 53.65 & 27.80 & 53.65 & 68.25
& 34.67 & 15.39 & 11.49 & 34.12 & 39.05 & 40.51
& 10.95 & 5.96  & 5.02  & 10.40 & 11.86 & 12.23 \\
& SWE-Agent
& 47.45 & \underline{47.45} & 47.45 & 70.07
& 36.86 & 38.75 & 37.90 & 35.58 & 58.12 & 61.19
& 30.66 & 29.99 & 25.91 & 28.22 & 44.59 & 47.26 \\
& OrcaLoca 
& \underline{65.69} & 40.08 & \underline{65.69} & 70.80
& 45.62 & 27.74 & 27.12 & 44.83 & 48.60 & 48.60
& \underline{43.07} & \underline{32.97} & \underline{32.41} & \underline{40.45} & \underline{51.89} & 52.25 \\
& MoatlessTools
& 51.82 & 46.78 & 51.82 & \underline{71.90}
& 42.34 & \underline{39.96} & \underline{40.29} & 41.24 & 61.68 & 64.96
& 33.58 & 32.36 & 31.70 & 30.93 & 51.19 & \underline{52.65} \\
\midrule
Procedure
& \tool
& \textbf{69.71} & \textbf{59.85} & \textbf{69.71} & \textbf{76.64}
& \textbf{60.22} & \textbf{47.75} & \textbf{47.29} & \textbf{57.30} & \textbf{68.19} & \underline{68.73}
& \textbf{48.18} & \textbf{34.37} & \textbf{33.49} & \textbf{41.74} & \textbf{52.93} & \textbf{53.66} \\
\bottomrule
\end{tabular}
\end{table*}

\tool achieves state-of-the-art performance across all granularities and metrics. 
Specifically, \tool with GPT-4o attains 70.07\% file-level Recall@1, 
outperforming the best baseline Agentless by 17.78\%. 
The improvements are more pronounced at finer granularities: 
\tool achieves 60.58\% / 46.72\% Precision@1 and 58.39\% / 41.07\% Recall@1 at module and function levels respectively, 
representing a notable boost of 25.98\% and 41.57\% in Recall@1 over the strongest baselines 
(Agentless at module level and MoatlessTools at function level); 
it further improves function-level Recall@3 by 23.38\% over the strongest baseline. 
Similar gains hold with Claude, indicating that \tool is robust across backbone models, as test-mediated localization (issue$\rightarrow$tests$\rightarrow$code) reduces reliance on direct issue--code matching.

\subsection{RQ2. Issue Resolution Performance}
\label{sec:downstream}

\begin{table}[h]
\centering
\small
\caption{RQ2. End-to-end issue resolution on SWE-bench Lite.}
\vspace{-5pt}
\label{tab:downstream}
\begin{tabular}{@{}lcc@{}}
\toprule
\textbf{Method} & \textbf{Resolved} & \textbf{Rate (\%)} \\
\midrule
Agentless (original) & 96 & 32.00 \\
Agentless + \tool & 113 (+17 $\uparrow$) & 37.67 (+5.67 $\uparrow$) \\
\bottomrule
\end{tabular}
\end{table}

To evaluate whether improved localization translates to better end-to-end performance, we integrate \tool into the Agentless pipeline~\citep{xia2024agentless}, replacing its original localization module while keeping the rest unchanged.

\autoref{tab:downstream} presents the results on SWE-bench Lite. 
\tool-augmented Agentless achieves a resolution rate of 37.67\%, compared to 32.00\% for the original Agentless, representing 17 additional resolved issues. 
This improvement demonstrates that superior @1 localization performance, particularly the advantages of fine-grained localization at the function levels, directly benefits downstream patch generation by providing the repair model with a highly focused and relevant context. 
Unlike agentic search, which must discover relevant tests and infer their connections to implementation through heuristic exploration, \tool explicitly decomposes localization into the issue$\rightarrow$tests$\rightarrow$code pathway and uses execution traces as structured intermediate evidence.
By pinpointing exact edit locations rather than providing entire files, \tool effectively avoids overwhelming the model with extraneous code, thereby reducing potential distractions during the reasoning process; detailed examples and cases illustrating this effect are provided on the artifact page~\cite{homepage}.

\subsection{RQ3. Cost Analysis}
\label{sec:efficiency}


We evaluate the cost-efficiency of \tool by measuring the average monetary cost per issue on SWE-bench Lite using GPT-4o, with results compared against representative baselines in \autoref{tab:efficiency}. Compared with representative agent-based methods, \tool achieves significant cost savings by reducing the average expense per issue by approximately 17.54\%, 41.98\%, and 46.89\% relative to OpenHands, SWE-Agent, and OrcaLoca, respectively, averaging 35.47\% across these baselines. Although \tool incurs a slightly higher cost than the procedural baseline Agentless due to the overhead of dynamic trace analysis, this modest investment is justified by its superior performance. \tool significantly outperforms Agentless in localization accuracy (\autoref{sec:main_results}) and enables the resolution of 17.72\% more issues in downstream tasks (\autoref{sec:downstream}). In summary, \tool strikes an optimal balance between resource consumption and diagnostic effectiveness, providing state-of-the-art capabilities with much higher cost-efficiency than complex agents.


\begin{table*}[ht]
\centering
\small

\begin{minipage}[t]{0.40\textwidth}
\centering
\caption{RQ3. Efficiency and cost comparison (per issue, GPT-4o). 
\textbf{Bold}: best; \underline{Underline}: second best.}
\label{tab:efficiency}
\begin{tabular}{@{}lc@{}}
\toprule
\textbf{Method}    & \textbf{Cost (\$)} \\
\midrule
Agentless    & \textbf{0.70} \\
OpenHands    & 1.14 \\
SWE-Agent    & 1.62 \\
OrcaLoca     & 1.77 \\
\midrule
\tool        & \underline{0.94} \\
\bottomrule
\end{tabular}
\end{minipage}
\hfill
\begin{minipage}[t]{0.55\textwidth}
\centering
\caption{RQ4. Ablation study on SWE-bench Lite.}
\label{tab:ablation}
\begin{tabular}{@{}lccc@{}}
\toprule
\multirow{2}{*}{\textbf{Configuration}} & \multicolumn{3}{c}{\textbf{F1@3 (\%)}} \\
\cmidrule(l){2-4}
& File & Module & Function \\
\midrule
\tool (full) & \textbf{65.27} & \textbf{51.07} & \textbf{34.08} \\
\midrule
w/o Tests & 52.55 & 41.27 & 21.14 \\
w/o Domain Knowledge & 58.47 & 43.47 & 26.38 \\
w/o Trace Analysis & 56.02 & 42.03 & 25.55 \\
w/o Refinement & 62.71 & 45.72 & 28.89 \\
Module Refinement & \underline{64.54} & \underline{47.92} & \underline{31.74} \\
w/o Reranking & 50.97 & 24.32 & 16.24 \\
\bottomrule
\end{tabular}
\end{minipage}

\end{table*}

\subsection{RQ4. Ablation Study}
\label{sec:ablation}


We conduct ablation experiments to understand the contribution of each component, reporting F1@3 scores across three granularities in Table~\ref{tab:ablation}.
Removing the test-based intermediate entirely (w/o Tests), i.e., replacing Relevant Test Retrieval and Trace-Guided Localization with direct retrieval, causes significant degradation, with F1@3 dropping by 12.72, 9.80, and 12.94 percentage points at the file, module, and function levels. This confirms that leveraging tests as structured intermediate evidence is central to \tool's effectiveness.
Removing trace analysis results in a performance decrease across all metrics, with an average drop of 8.94\% and specific declines of 9.25\%, 9.04\%, and 8.53\% at the file, module, and function levels. These results emphasize that execution-based call hierarchies are essential for establishing causal links between tests and code, effectively mitigating the noise inherent in static analysis. Excluding domain knowledge leads to an average 7.37\% reduction in performance. This component facilitates the alignment of issue requirements with project-specific test identifiers by leveraging historical commit data to bridge terminology gaps. 
The removal of the refinement stage produces a smaller average decline of 4.37\%, suggesting that in the majority, the \tool achieves robust localization performance within the space covered by existing tests alone. 
Replacing the full refinement process with module-level refinement limits the search space while preserving performance near the full \tool configuration. This observation suggests that when operating under restricted budgets, narrowing the refinement scope can further improve system efficiency with minimal impact on localization accuracy. 
The substantial performance loss in the absence of reranking is attributed to the F1@3 experimental setting, as this module provides more accurate importance-based ordering of entities. Furthermore, prioritizing the most probable edit locations at the top of the recommendation list reflects a method design that aligns with the requirements of real-world software maintenance tasks.

\section{Failure Analysis}

While \tool achieves promising performance in issue localization, it still fails to identify the correct edit locations in several recurring scenarios. This section focuses specifically on these failure modes. For each failure mode, we qualitatively analyze the underlying design choice or assumption that causes the failure, illustrate it with a representative example, and discuss potential mitigation directions for future work.

\subsection{Localization Failure due to Missing Test Coverage}

\label{sec:failure_missing_coverage}

While \tool's test-driven localization paradigm achieves notable effectiveness in bridging issue descriptions with target code locations, a specific failure mode arises when the repository's existing test suite fails to adequately cover the buggy code regions. In such scenarios, the hierarchical trace graph $G_t$ constructed from retrieved test execution cannot propagate localization signals to uncovered code regions, manifesting as $\text{Cov}(\mathcal{T}_d) \cap \mathcal{G} = \emptyset$, where the union of dynamic coverage from all retrieved tests $\mathcal{T}_d$ exhibits no intersection with the ground-truth edit set $\mathcal{G}$. This represents a boundary condition for the H1 (Coverage Feasibility) hypothesis established in~\autoref{sec:empirical}. Even after filtering for repositories with at least 50\% test function execution success (\autoref{sec:empirical_setup}), incomplete coverage persists as a structural challenge. This reflects the varying test maturity levels found in real-world codebases, an inherent constraint of the test-driven localization paradigm when applied at scale.

\begin{figure*}[htbp]
    \centering
    \includegraphics[width=\textwidth]{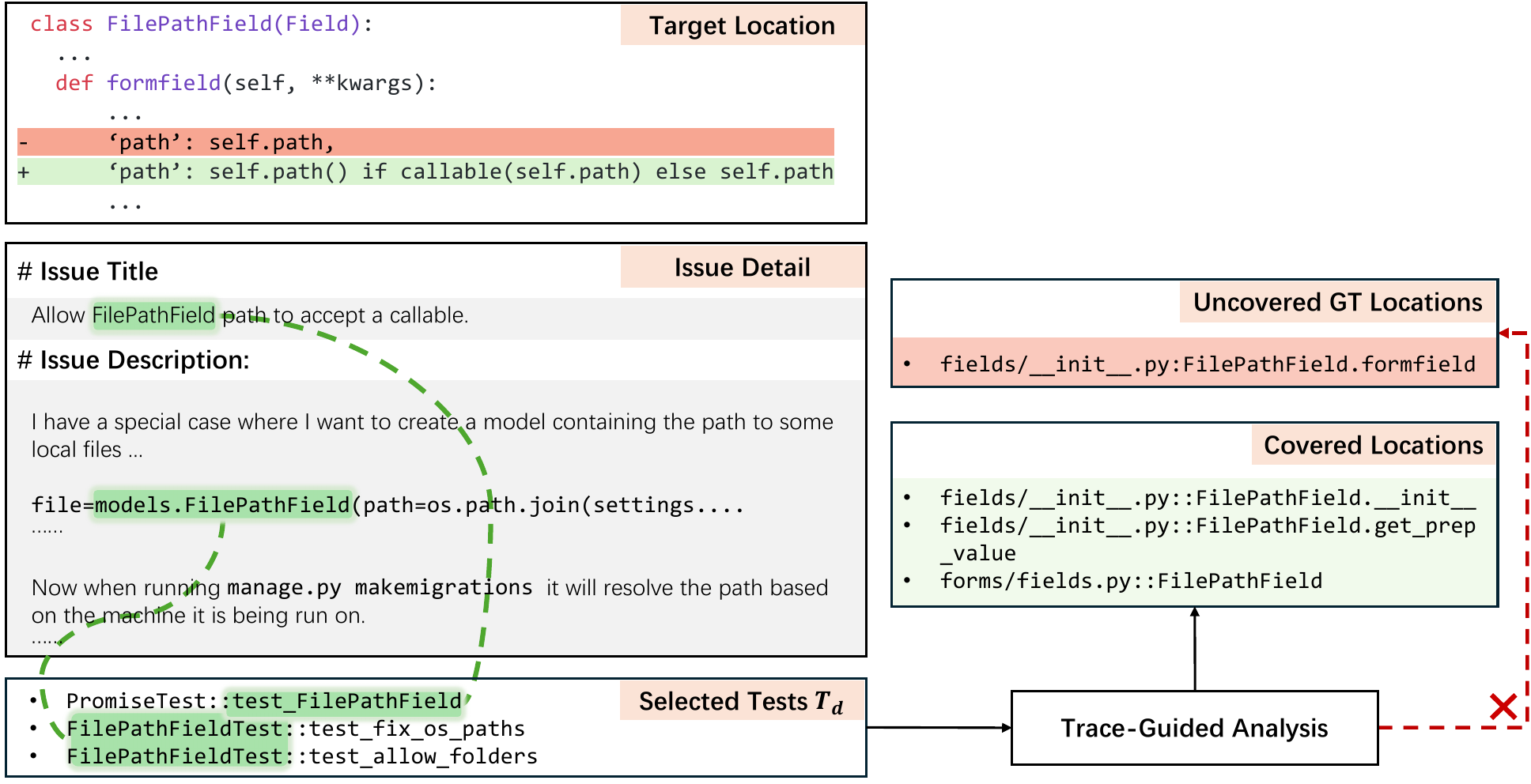}
    \caption{
        Illustration of the coverage gap failure mode on \texttt{django\_\_django-10924}.
        The selected tests $\mathcal{T}_d$ cover auxiliary methods of the target class but
        do not execute the ground-truth edit location (shown in red), causing it to be
        excluded from the trace-guided suspicious set $S$ throughout all pipeline stages.
    }
    \label{fig:bad_case_coverage}
\end{figure*}

Specifically, as depicted in~\autoref{fig:bad_case_coverage}, although the retrieval stage surfaces tests that are semantically related to the target functionality, the selected tests $\mathcal{T}_d$ collectively cover only auxiliary or adjacent locations, leaving the ground-truth edit method entirely absent from $\text{Cov}(\mathcal{T}_d)$. Consequently, the trace-guided analysis can only reason over the covered locations and receives no direct execution evidence for the uncovered ground-truth method, causing it to be excluded from the suspicious set and the final localization output. This case thus exemplifies how semantic relevance at the test-retrieval stage does not guarantee execution reachability at the method level, a distinction that is invisible to retrieval-only approaches but becomes a hard boundary when trace-guided localization is constrained by incomplete coverage.

The scope of this limitation is empirically grounded in the H1 analysis of~\autoref{sec:empirical}, which establishes that 33.30\% of ground-truth functions fall outside the full test suite's coverage, indicating that coverage gaps are a non-negligible structural risk in real-world instances. Critically, this boundary cannot be overcome by improving the semantic alignment between issues and tests (H2) or by strengthening the two-hop bridging pathway (H3), since all downstream stages are fundamentally constrained by what the selected tests execute at runtime. A natural direction for future work is to combine \tool with issue-to-test generation techniques~\citep{nashid2025issue2test} that synthesize bug-triggering tests directly from issue descriptions, thereby relaxing the dependency on pre-existing test coverage and extending the reach of trace-guided localization to repositories with lower test maturity.

\subsection{Localization Failure under Long Call Chains}

\label{sec:failure_deep_call_chain}

Even when the ground-truth edit location \emph{is} covered by the retrieved tests' execution traces, localization can still fail when the ground-truth node is reachable only through a deep and complex call chain. The execution path from the selected tests to the ground-truth location passes through a large number of intermediate orchestration layers, whose high invocation frequency causes them to dominate the trace graph $G_t$ and obscure the true fix site. As a result, the localization pipeline fails to recover $\mathcal{G}$ in $L^*$ not due to an absence of execution evidence, but because the signal-to-noise ratio within $G_t$ degrades with increasing chain depth, rendering the ground-truth node statistically indistinguishable from higher-level infrastructure components. This constitutes a qualitatively distinct failure mode from the coverage gap discussed in~\autoref{sec:failure_missing_coverage}.

\begin{figure*}[htbp]
    \centering
    \includegraphics[width=\textwidth]{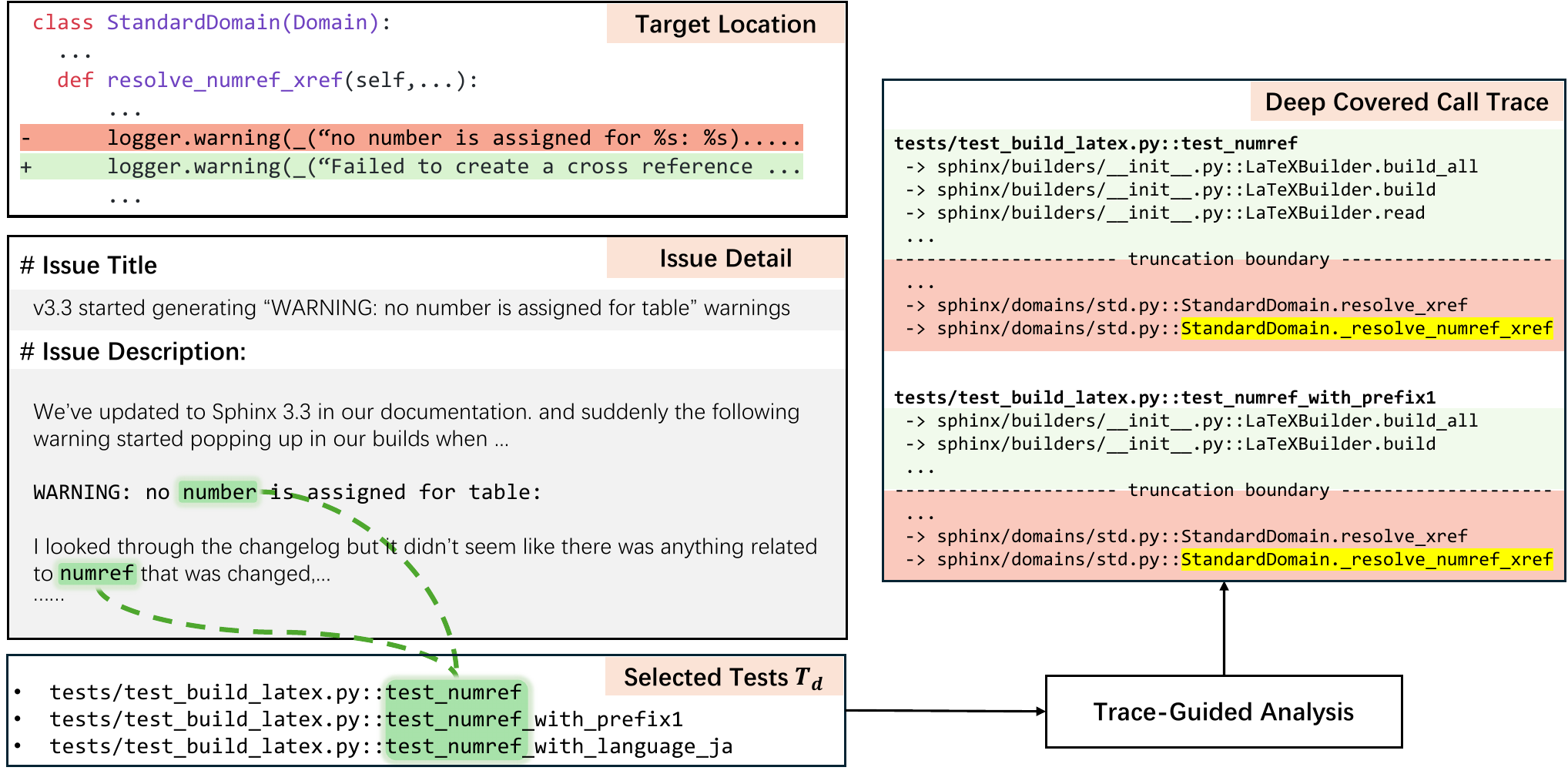}
    \caption{
        Illustration of the deep call chain failure mode on \texttt{sphinx-doc\_\_sphinx-8474}.
        Although the ground-truth edit location (shown in red) is reachable within
        $\text{Cov}(\mathcal{T}_d)$, it is buried beneath multiple layers of orchestration
        nodes that dominate the trace-guided suspicious set $\mathcal{S}$, preventing its
        recovery as a final candidate $L^*$. The truncation boundary marks the depth limit imposed by BFS-based trace pruning; nodes below this boundary, including the ground-truth location, are omitted from the trace provided to the trace-guided analysis stage.
    }
    \label{fig:bad_case_chain}
\end{figure*}

As illustrated in~\autoref{fig:bad_case_chain}, the selected tests are semantically well-aligned with the issue and their execution traces do reach the ground-truth edit location through the full build pipeline. However, the trace is dominated by high-level orchestration components that are invoked far more frequently and appear structurally more prominent within $G_t$. As a result, the contextual refinement stage retrieves class-level context around these upstream nodes, and the subsequent reranking elevates pipeline-level candidates over the precise domain-level resolver where the actual fix resides. This exposes a characteristic difficulty in trace-guided localization, where high-level pipeline components are both highly visible in the trace and semantically plausible given the issue description, making it difficult to distinguish them from the true root-cause location without reasoning over the depth and role of the covered call path.

The prevalence of this failure mode is closely tied to the architectural complexity of the target codebase. Projects with deeply layered execution pipelines, such as build systems, compilers, or documentation frameworks, are particularly susceptible, as their runtime behavior is inherently dominated by high-traffic orchestration components that sit far above the actual logic being modified. A promising mitigation direction is to incorporate call-chain depth as an explicit signal during candidate ranking, discounting locations that appear primarily as transitive callers rather than direct implementors of the relevant behavior. Complementarily, future work could leverage static call graph analysis to prune orchestration-layer noise from the trace prior to suspicious set construction, thereby improving the ability to surface deep but precise edit locations.

\section{Threats to Validity}

\textbf{Internal Validity}
The effectiveness of \tool's domain knowledge enhancement depends on the availability and quality of historical commit data.
If a repository has insufficient commit history or maintains messy, non-atomic commit logs, the AST-based mining process may fail to extract meaningful domain tokens, potentially weakening the semantic alignment between issues and tests.
While our framework implements filtering mechanisms to exclude low-quality commits, the reliance on historical signals remains a potential internal threat.
Future work could involve incorporating external documentation or API references to supplement missing local history.


\noindent\textbf{External Validity}
A primary external threat is the generalizability of \tool to issues that cannot be connected to runnable tests or partially executable behaviors.
\tool is designed for the runnable and test-accessible issue localization setting, where tests can serve as executable proxies for requirements.
For non-runnable issues, or cases where the issue behavior is not exercised by existing tests, the ground-truth edit locations may not be captured within execution traces.
However, our empirical study suggests that even in large-scale repositories, existing tests often exercise core functional paths, providing a high-recall starting point.
Future work could combine \tool with issue-to-test generation techniques to synthesize bug-triggering tests from issue descriptions, thereby relaxing the dependency on existing test coverage.

Another external threat is language and framework support.
Currently, our implementation and evaluation focus exclusively on Python repositories and the \texttt{sys.settrace} instrumentation.
The characteristics of dynamic tracing and caller-callee relationship extraction may differ in statically typed languages like Java or C++.
While the core methodology of test-mediated localization is language-agnostic, the current lack of cross-language validation remains a threat to external validity.
Generalizing \tool to other languages mainly requires adapting language-specific components, including test collection, execution tracing, and the language frontend, while the core issue-to-test-to-code pipeline remains unchanged.
We plan to implement support for additional programming languages in future iterations.

\noindent\textbf{Construct Validity}
The process of collecting full execution traces and constructing trace graphs can be computationally expensive for massive software systems.
This overhead might raise concerns regarding the practical utility of the tool in rapid CI/CD environments.
To mitigate this threat, we emphasize that trace collection is primarily an offline preprocessing step.
Furthermore, this process can be optimized through incremental trace updates where only affected tests are re-executed and traced.
Such engineering optimizations will ensure that the system remains scalable as the project evolves.

\section{Related Work}


\textbf{Information Retrieval and Bug Localization.}
Retrieval-based approaches compute similarity between issues and code to bridge the gap between natural language and formal syntax~\citep{Zhou2012Where, Antoniol2002Recovering}.
Early works relied on sparse retrievers like BM25~\citep{robertson2009probabilistic} utilizing lexical overlap.
To improve accuracy, subsequent studies introduced query reformulation with contextual cues~\citep{rahman2018improving} or composed richer evidence sources such as code change histories, metadata, and project-specific edit patterns~\citep{youm2017improved, wang2016amalgam+, liu2024coedpilot, liu2025learning}.
While effective, these enhancements introduce sensitivity to modeling choices, where configuration substantially impacts localization performance~\citep{tantithamthavorn2018impact}.
More recently, dense retrievers such as CodeSage~\citep{zhang2024code} and CodeRankEmbed~\citep{suresh2024cornstack} have emerged to capture semantic relationships beyond keyword matching.
This paradigm has been further adapted to specialized scenarios: CoRet~\citep{fehr2025coret} incorporates call graph dependencies, BLAZE~\citep{chakraborty2025blaze} addresses cross-project settings, and other models target concurrent programs~\citep{shao2023information}.
However, as noted in classic studies, IR-based methods remain largely static proxies for relevance and continue to struggle with the "vocabulary mismatch" problem where informal issue descriptions and formal code syntax lack sufficient overlap~\citep{Marcus2003Recovering}.

\noindent\textbf{Traditional Fault Localization and Traceability.}
Our work is fundamentally grounded in the long-standing challenges of Spectrum-based Fault Localization (SBFL) and Traceability Link Recovery (TLR).
SBFL techniques, such as Tarantula~\citep{jones2005empirical} and Ochiai~\citep{abreu2006evaluation}, utilize program spectra from passing and failing tests to rank suspicious code entities.
While foundational, empirical studies show that SBFL effectiveness varies significantly across scales and often requires combination with additional signals~\citep{de2016spectrum, keller2017critical, jiang2019combining}.
Crucially, these methods rely on pre-existing failing tests, limiting their applicability in general issue resolution where such tests are often absent. Complementary research has explored project-specific test generation for requirement validation and test generalization for broader scenario coverage~\citep{qi2018generating, qi2026generalizing}.
Traceability Link Recovery (TLR) aims to link requirements to code but remains costly to maintain and recover automatically~\citep{Antoniol2002Recovering, aung2020literature}.
Related research in Feature Location~\citep{dit2013feature} and dynamic analysis has demonstrated that execution scenarios and traces can provide direct evidence of functional behavior~\citep{eisenbarth2001aiding, salah2006scenario}.
However, interpreting these dynamic traces has historically relied on manual effort or complex static analysis.
Bridging this gap by automating diagnostic reasoning on execution traces—potentially via LLMs—remains an open challenge for recovering dynamic traceability links at scale.

\noindent\textbf{Procedure-based Approaches} employ hierarchical pipelines to narrow the search space. Agentless~\citep{xia2024agentless} pioneered a three-phase paradigm, while SWE-Fixer~\citep{xie2025swe} streamlines this with BM25 retrieval and reranking. BugCerberus~\citep{chang2025bridging} extends to statement-level localization with program slicing, and PatchPilot~\citep{li2025patchpilot} adds reproduction and refinement stages. Related work also uses formal refinement to guide and verify LLM-generated code~\citep{cai2025automated}. While efficient, these rigid pipelines are prone to irreversible error propagation, particularly when the initial retrieval step relies on suboptimal queries or configurations~\citep{rahman2018improving, tantithamthavorn2018impact}.

\noindent\textbf{Agent-based Approaches} frame localization as sequential decision-making. SWE-Agent~\citep{yang2024swe} designs an Agent-Computer Interface for navigation, while OpenHands~\citep{wang2024openhands} provides Docker-sandboxed environments. Multi-agent systems like MASAI~\citep{arora2024masai} and SWE-Search~\citep{antoniades2024swe} distribute tasks across sub-agents. Graph-based methods have also emerged: LocAgent~\citep{chen2025locagent} constructs heterogeneous graphs with multiple edge types; RepoGraph~\citep{ouyang2024repograph} supports k-hop retrieval; CodexGraph~\citep{liu2025codexgraph} enables Cypher-based querying. Despite their sophistication, these methods navigate \textit{structural} rather than \textit{functional} relationships, missing disconnected targets.

\section{Conclusion}
This paper revisits issue localization from a requirement-centric perspective and argues that 
\emph{test suites can serve as executable requirements} that bridge the abstraction gap between natural-language issues and code.
Motivated by both theoretical analysis and large-scale empirical evidence,
we propose \tool, a test-driven localization framework that leverages domain-knowledge-enhanced test representations for 
effective issue--test alignment and hierarchical execution trace analysis to
filter infrastructure noise and pinpoint requirement-central code.

Extensive experiments on SWE-bench Lite demonstrate
that \tool achieves state-of-the-art localization performance
across file, module, and function levels.
When integrated into an end-to-end issue resolution pipeline,
the improved localization quality translates into tangible gains in
downstream patch generation, confirming that precise, requirement-aware
localization is a critical enabler for automated software maintenance.
Our in-depth analysis further reveals that the advantages of test-driven
localization are most pronounced for complex, multi-file issues,
while also highlighting limitations arising from incomplete test coverage
and noisy execution traces.

\section*{Data Availability}
The source code of \tool, LLM prompts, detailed analysis from the theoretical motivation and empirical study, illustrative case studies, and all other supplementary materials are available at \cite{homepage}.

\begin{acks}

This research is supported in part by the National Natural Science Foundation of China (62572300), the Minister of Education, Singapore (MOE-T2EP20124-0017, MOET32020-0004), the National Research Foundation, Singapore and the Cyber Security Agency under its National Cybersecurity R\&D Programme (NCRP25-P04-TAICeN), DSO National Laboratories under the AI Singapore Programme (AISG Award No: AISG2-GC-2023-008-1B), and Cyber Security Agency of Singapore under its National Cybersecurity R\&D Programme and CyberSG R\&D Cyber Research Programme Office, and partially by HUAWEI’s Al Hundred Schools Program using the Ascend AI technology stack. Any opinions, findings and conclusions or recommendations expressed in this material are those of the author(s) and do not reflect the views of National Research Foundation, Singapore, Cyber Security Agency of Singapore as well as CyberSG R\&D Programme Office, Singapore.

\end{acks}

\bibliographystyle{ACM-Reference-Format}
\bibliography{example_paper}
\end{document}